\documentclass[prl,showpacs,amsmath,assymb,eufrak]{revtex4}
\usepackage{bm}
\usepackage{epsfig,amsfonts,amsbsy}

\newcommand{\sbkt}[1]{\langle#1\rangle}
\newcommand{\bbkt}[1]{\bigl\langle#1\bigr\rangle}
\newcommand{\Bbkt}[1]{\Bigl\langle#1\Bigr\rangle}

\newcommand{\WT}{\mathcal{T}}
\newcommand{\calD}{\mathcal{D}}

\newcommand{\pathG}{\hat{\Gamma}}

\newcommand{\patha}{\hat{\bm \alpha}}
\newcommand{\pathad}{\hat{\bm \alpha}^\dagger}

\newcommand{\rhost}{\rho}
\newcommand{\rhoeq}{\rho_\mathrm{eq}}

\newcommand{\oet}{O(\varepsilon^2)}

\newcommand{\Di}{\mathit{\Delta}}
\newcommand{\Da}{\Di\bm\alpha}

\newcommand{\ti}{0}
\newcommand{\tf}{\tau}

\newcommand{\intt}{\int_{\ti}^{\tf} dt}

\newcommand{\eq}{\mathrm{eq}}

\begin{document}
\title{
Universal expression for adiabatic pumping in terms of nonequilibrium steady states
}
\author{Naoko Nakagawa}
\affiliation {
College of Science,  Ibaraki University, Mito, Ibaraki 310-8512, Japan
}

\date{\today}

\begin{abstract}
We develop a unified treatment of pumping and nonequilibrium thermodynamics.
We show that the pumping current generated through an adiabatic mechanical
operation in equilibrium
can be expressed in terms of the stationary distribution of the
corresponding driven nonequilibrium system.
We also show that the total transfer in pumping can be evaluated from
the work imported to the driven counterpart.
These findings lead us to a unified viewpoint for pumping
and nonequilibrium thermodynamics. 
\end{abstract}


\maketitle


For centuries, heat pumping has been considered an important topic.
The Carnot engine showed the direct relation between mechanical work and pumping of heat.
Pumping induced by electric current known as the Peltier effect, was explained by
the linear response theory as an example of the reciprocal relation.
In molecular scales, the possibility of realizing heat pumps 
with thermal ratchets is suggested in
\cite{Ai_Wang_Liu,Segal_Nitzan,Broeck_Kawai,NK-pump}

Apart from heat pumps, ion pumps or the directive transport of biomolecules
are theoretically intensively studied.
These are modeled using flashing ratchets \cite{Magnasco,Derenyi_Vicsek,Julicher_Ajdari_Prost,Astumian-ratchet,Reimann} 
as stochastic pumps in molecular scales.
The mechanism of pumping in flashing ratchets 
is related to 
geometric effects in the parameter space
\cite{Parrondo,Sinitsyn_Nemenman,Astumian,Ohkubo}.
The same property in heat pumps has also been discussed \cite{Ren_Hanggi_Li}.
These studies suggest that the universal characteristics of pumps 
exist in various designs.

In this paper,
we develop a unified viewpoint on pumping and nonequilibrium thermodynamics, from which
one can derive the universal characteristics of pumps
as well as examine efficient protocols for pumping.
By pumping, we mean  an equilibrium process in which 
the parameters of the system 
are varied through an external agent
according to a fixed protocol 
in order to invoke the desired type of current through the system.
For each setup of pumping, 
we introduce a corresponding ``{\it driven counterpart}'',
i.e., a nonequilibrium system 
in which the current flows spontaneously owing to an applied driving field.
We then show that the pumping current 
is expressed in terms of the stationary probability distribution 
of the driven system
and that it is well evaluated from the work imported to the driven counterpart operated using the same protocol.

\section{Setup}

We employ a classical system with a Hamiltonian $H(\Gamma)$,
where $\Gamma=(\{\bm{x}\},\{\bm{p}\})$ denotes the system's microstate.
The Hamiltonian depends on a set of parameters $\bm\alpha=(\alpha_1,\alpha_2,\cdots,\alpha_n)$. 
We assume a time-reversal symmetry for the Hamiltonian
$H(\Gamma)=H(\Gamma^*)$, 
where $\Gamma^*=(\{\bm{x}\},\{-\bm{p}\})$.
We do not limit the number of the system's degrees of freedom.
It may be one or the Avogadro number.
The system is not isolated but is in contact with an equilibrium environment (baths).

The time evolution of the system is governed by the deterministic dynamics according to
the Hamiltonian $H(\Gamma)$ and
the stochastic Markovian dynamics owing to the external bath coupling.
One operates the system mechanically by varying the parameters $\bm\alpha$.
The protocol for this operation is denoted as
$\patha :=(\bm\alpha(t))_{t\in[\ti,\tf]}$.
When discussing the time evolution of $\Gamma$,
we denote its value at time $t$ by $\Gamma(t)$ 
and 
its path in the whole time interval $[\ti,\tf]$
by $\pathG=(\Gamma(t))_{t\in[\ti,\tf]}$.

In order to theoretically analyze pumping problems, 
we also study a system driven by a certain driving field $\varepsilon$.
We assume that the system reaches a unique nonequilibrium steady state (NESS)
when we fix $\varepsilon$ and $\bm\alpha$ for a sufficiently long time.
The transition probability 
 associated with the path $\pathG$
is denoted by $\WT_{\patha,\varepsilon}(\pathG)$
 in a protocol $\patha$ under the driving $\varepsilon$.
The probability distribution in the unique NESS is denoted by $\rhost_{\varepsilon}(\Gamma)$, with which we define
\begin{equation}
\psi^{\varepsilon}(\Gamma):=-\log \rhost_{\varepsilon}(\Gamma)
.
\end{equation}
Note that $\rhost_{\varepsilon}(\Gamma)$ depends on $\bm\alpha$
although we do not specify it for simplicity of notation.
The canonical distribution $\rhoeq(\Gamma)$ corresponds to $\rhost_{0}(\Gamma)$,
and we use $\psi^\eq$ instead of $\psi^{0}$.

For any function $f(\pathG)$ of a path, 
we define its average in the protocol $\patha$
as 
\begin{equation}
\bbkt{f}_{\varepsilon}:=\int{\calD}\pathG\rhost_{\varepsilon}(\Gamma(\ti))\WT_{\patha,\varepsilon}(\pathG)f(\pathG),
\end{equation}
where $\int{\calD}\pathG(\cdots)$ denotes the integral 
over all the possible paths $\pathG$.
For any function $f(\Gamma)$ of a state, we define its average in the steady state
as 
\begin{equation}
\bbkt{f}_{\rho_\varepsilon}:=\int d\Gamma \rhost_{\varepsilon}(\Gamma)f(\Gamma).
\end{equation}
For equilibrium processes ($\varepsilon=0$), we use $\sbkt{f}_\eq$ and $\sbkt{f}_{\rhoeq}$
instead of $\sbkt{f}_{0}$ and $\sbkt{f}_{\rho_0}$, respectively.

We assume that the current at time $t$
is the function of $\pathG$,  $J(\pathG;t)$,
i.e. it depends only on the system's path but not on the system's environment.
Because the probability of the path depends on the environment
and the applied protocol,
the average $\sbkt{J}_{\varepsilon}$ in turn depends on them.
The total transfer in the whole time interval is given by
\begin{equation}
Q(\pathG)=\intt J(\pathG;t)
.
\end{equation}
In the context of a pump,
$Q(\pathG)$ is the ``{\it total pumping}'' in a single execution of the protocol.

\section{Pumping current  and its conjugate driving}
For a heat pump carrying energy from one place to the other, 
$J(\pathG;t)$ is the heat current between the two places
and $Q(\pathG)$ is the total transferred heat.

It is crucial for us to observe that the mean heat current can be produced not only by 
the mechanical operation for pumping but also by imposing 
a difference in the temperatures at the two places.
In the latter case, the mean current flows spontaneously along the natural direction,
satisfying the second law of thermodynamics.
The difference of the inverse temperatures is often called thermodynamic force
corresponding to the heat current. 
In this paper, we call it  the {\it conjugate driving} corresponding to the heat current.

We refine the above situation as follows:
In order to study the heat pumping in the system 
in contact with two separate isothermal heat baths indexed by $k$ ($k=1,2$),
for which the inverse temperature is denoted by $\beta$,
we also study its counterpart with the conjugate driving, 
i.e. the same system,
for which the inverse temperatures $\beta_1$, $\beta_2$ of the baths 
are different.
We choose $\beta_k$ so as to satisfy $\beta=(\beta_1+\beta_2)/2$.

Letting $J_k(\pathG;t)$ be the heat current from the $k$th heat bath to the system at time $t$ in the path $\pathG$,
the heat current from one heat bath to the other is formulated as
\begin{equation}
J(\pathG;t)=\frac{J_1(\pathG;t)-J_2(\pathG;t)}{2}
\label{e:J}
\end{equation}
for both the pumping system and its driven counterpart.
By computing the average, we have $\sbkt{J}=\sbkt{J_1}=-\sbkt{J_2}$ 
under steady driving or any cyclic protocol.
The conjugate driving, i.e. the thermodynamic force corresponding to the heat current
is
\begin{equation}
\varepsilon=\beta_2-\beta_1.
\end{equation}
The entropy production owing to the heat current is $\varepsilon J(\pathG;t)=(\beta_2-\beta_1)J(\pathG;t)$.

For stochastic pumps represented using flashing ratchet models (see Fig.~\ref{fig:ratchet}),
we consider a particle in a potential with a periodic boundary condition in a certain coordinate $x$.
When applying a cyclic operation to the potential, the system may have a nonvanishing circulation in its microstates, 
and this may be observed as directed mean current $\sbkt{J}_\eq$ of the particle,
where
\begin{equation}
J(\pathG;t)=\dot x(t).
\label{e:J2}
\end{equation}
We notice that $J(\pathG;t)$ is determined by the system's microstate 
and not by the operation.

The driven counterpart is the same system in which the particle is pulled
by a constant nonconservative force $f$ along the coordinate $x$.
The conjugate driving is
\begin{equation}
\varepsilon=\beta f,
\end{equation}
and the entropy production is $\beta f J(\pathG;t)$,
where $J(\pathG;t)$  for the driven system is the same as Eq.~\eqref{e:J2}.

Even though we present our claims for a closed system setup in this paper, 
they can also be extended to include open systems with particle baths 
by modifying the setup, as discussed in Sec.~5 of \cite{KN-long}.
For such open systems, we can consider particle pumping between two particle baths,
where the particle current is defined parallel to Eq.~\eqref{e:J}.
Here, the conjugate driving corresponds to
$\varepsilon=\beta(\mu_2-\mu_1)$,
where $\mu_k$ is the chemical potential for the $k$th particle bath.

\section{Main results}

\subsection{The expression for total pumping}

Our main result is the expression for the total pumping
produced in equilibrium adiabatic operations,
the derivation for which is given in the Appendix.

For the adiabatic protocol $\patha$ applied to an equilibrium system,
the total pumping is 
\begin{eqnarray}
\bbkt{Q}_\eq
&=&
\int_{\patha} d{\bm \alpha}\cdot
\Bbkt{
\left.
\nabla_{\bm\alpha}\psi^\eq
~\partial_\varepsilon\psi^{\varepsilon}
\right|_{\varepsilon=0}
}_{\rhoeq}
\label{e:pump1}
\\
&=&
\int_{\patha} d{\bm \alpha}\cdot
\Bbkt{
\nabla_{\bm\alpha}
\left.
\partial_\varepsilon
\psi^{\varepsilon}
\right|_{\varepsilon=0}
}_{\rhoeq}
,
\label{e:pump1-2}
\end{eqnarray}
where $\int_{\patha}d{\bm\alpha} \cdots$ is the line integral 
along the protocol $\patha$ in the parameter space of $\bm \alpha$.
It is remarkable that the total pumping is directly related to the 
steady probability distribution $\rhost_{\varepsilon}(\Gamma)$ for the 
driving counterpart.
$\rhost_{\varepsilon}(\Gamma)$
depends on
the type of the conjugate driving $\varepsilon$,
as does
the equilibrium pumping.

The expression \eqref{e:pump1} indicates that the pumping is efficient 
when $\nabla_{\bm\alpha}\psi^\eq(\Gamma)$ is parallel 
to $\partial_{\varepsilon}\psi^{\varepsilon}(\Gamma)$ in the phase space of $\Gamma$.
In other words, it is efficient when the operation $\patha$ 
well mimics the nonequilibrium driving.
It is worth noting that the kernels of Eqs.~\eqref{e:pump1} and \eqref{e:pump1-2} 
correspond to the off-diagonal components of the Fisher information matrix
because
$
\sbkt{
\nabla_{\bm\alpha}
\partial_\varepsilon
\psi^{\varepsilon}|_{\varepsilon=0}
}_{\rhoeq}
=
\sbkt{
\nabla_{\bm\alpha}
\partial_\varepsilon
\psi^{\varepsilon}
}_{\rho_\varepsilon}
$
for $\varepsilon\rightarrow 0$.

In cyclic protocols $\patha_\mathrm{cyc}$, we can apply the Stokes' theorem
to $\oint d{\bm\alpha}\cdot\rhoeq(\Gamma)\nabla_{\bm\alpha}\psi^{\varepsilon}$ 
in the right-hand side of Eq.~\eqref{e:pump1-2}.
Therefore,
\begin{equation}
\bbkt{Q}_\eq=
\int_{S} dS~
\bbkt{
{\cal J}
}_{\rhoeq}
,
\label{e:pump3}
\end{equation}
where $S$ is the region in the parameter space
enclosed by the closed line of $\patha_\mathrm{cyc}$.
We call $\sbkt{{\cal J}}_{\rhoeq}$  the {\it pumping density}.
When the number of parameters is two, i.e., $\bm\alpha=(\alpha_1,\alpha_2)$, 
the pumping density is
\begin{equation}
{\cal J}(\Gamma)=
\left.
\partial_\varepsilon
\left[
\partial_{\alpha_1}\psi^{\varepsilon}(\Gamma)
\partial_{\alpha_2}\psi^\eq(\Gamma)
-
\partial_{\alpha_1}\psi^\eq(\Gamma)
\partial_{\alpha_2}\psi^{\varepsilon}(\Gamma)
\right]
\right|_{\varepsilon=0}
.
\label{e:density}
\end{equation}
Various studies relating pumping to a geometric effect or the Berry phase 
\cite{Parrondo,Sinitsyn_Nemenman,Astumian,Ohkubo}
report a result similar to Eq.~\eqref{e:density},
which is derived from the master equation or the cumulant generating function
in cyclic operations in equilibrium.
We emphasize that the key point of our formula \eqref{e:density} is 
the use of the probability distribution $\rhost_{\varepsilon}$ for the driven counterpart.

\subsection{Equilibrium pumping and work in a driven counterpart}

To apply Eqs.~\eqref{e:pump1}, \eqref{e:pump1-2} or \eqref{e:density} to the pumping problem, 
we need to determine the probability distribution $\rhost_{\varepsilon}(\Gamma)$.
Since $\rhost_{\varepsilon}(\Gamma)$ is not known in general, we show how
$\sbkt{Q}_\eq$ and $\sbkt{{\cal J}}_{\rhoeq}$ can be approximately evaluated from 
an observable quantity.

We apply an adiabatic cyclic operation to both the equilibrium system and its driven counterpart.
Then from Eq.~\eqref{e:pump1-2} we get an approximate equality,
\begin{equation}
\bbkt{Q}_\eq=
-\beta\frac{\bbkt{W}_{\varepsilon}}{\varepsilon~~}+O(\varepsilon)
,
\label{e:work2}
\end{equation}
which relates the quantities of these distinct systems.
The derivation of Eq.~\eqref{e:work2} is shown in the Appendix.
Thus, we can evaluate ``pumping in equilibrium''
from the measurement of ``work in a driven counterpart.''
Here, the work is given by
$\sbkt{W}_{\varepsilon}=
\int_{\patha} d{\bm\alpha} \cdot \sbkt{\nabla_{\alpha} H}_{\rho_\varepsilon}$.
The relation \eqref{e:work2} is consistent with the extended Clausius
equality in \cite{KNST,KNST-nl}.

For general cyclic operations with a finite speed,
the total pumping is related to work in the nonequilibrium counterpart as
\begin{equation}
\bbkt{Q}_\eq
=
-\frac{1}{\varepsilon}\log\bbkt{
e^{-\beta W}
}^{\dagger}_{\varepsilon}
+O(\varepsilon),
\label{e:workG}
\end{equation}
where $\sbkt{\cdot}^{\dagger}$ is the average along the reverse cyclic protocol $\pathad$, i.e. $\pathad=(\bm\alpha(\tf-t))_{t\in[\ti,\tf]}$.
The relation \eqref{e:workG} follows from 
an extended Jarzynski equality to NESS \cite{Nakagawa1}.

Relations \eqref{e:work2} and \eqref{e:workG} suggest a new approach to study 
pumping
when $Q$ is difficult to observe but $W$ is measurable.
Depending on the protocol,  relation  \eqref{e:work2} or \eqref{e:workG} may be useful.
Note that the Jarzynski-like form \eqref{e:workG} is more useful in mesoscopic pumps 
because the Jarzynski equality  \cite{Jarzynski} is known to be efficient in mesoscopic systems.

We expect that the map of $\sbkt{{\cal J}}_{\rhoeq}$ in the space of $\bm\alpha$
can be a powerful tool to design an efficient protocol for pumping.
Equation \eqref{e:pump3} indicates that
$\sbkt{{\cal J}}_{\rhoeq}$ is approximated by $\sbkt{Q}_\eq/S\simeq
-\beta\sbkt{W}_{\varepsilon}/(\varepsilon S)$,
when we apply a cyclic protocol with a sufficiently small area $S$ in the parameter space.

\section{Examples}

\subsection{Numerical demonstration  for stochastic pumping}

\begin{figure}[t]
\centering
\begin{picture}(190,150)
\put(-25,80){\large (a)}
\includegraphics[scale=0.4]{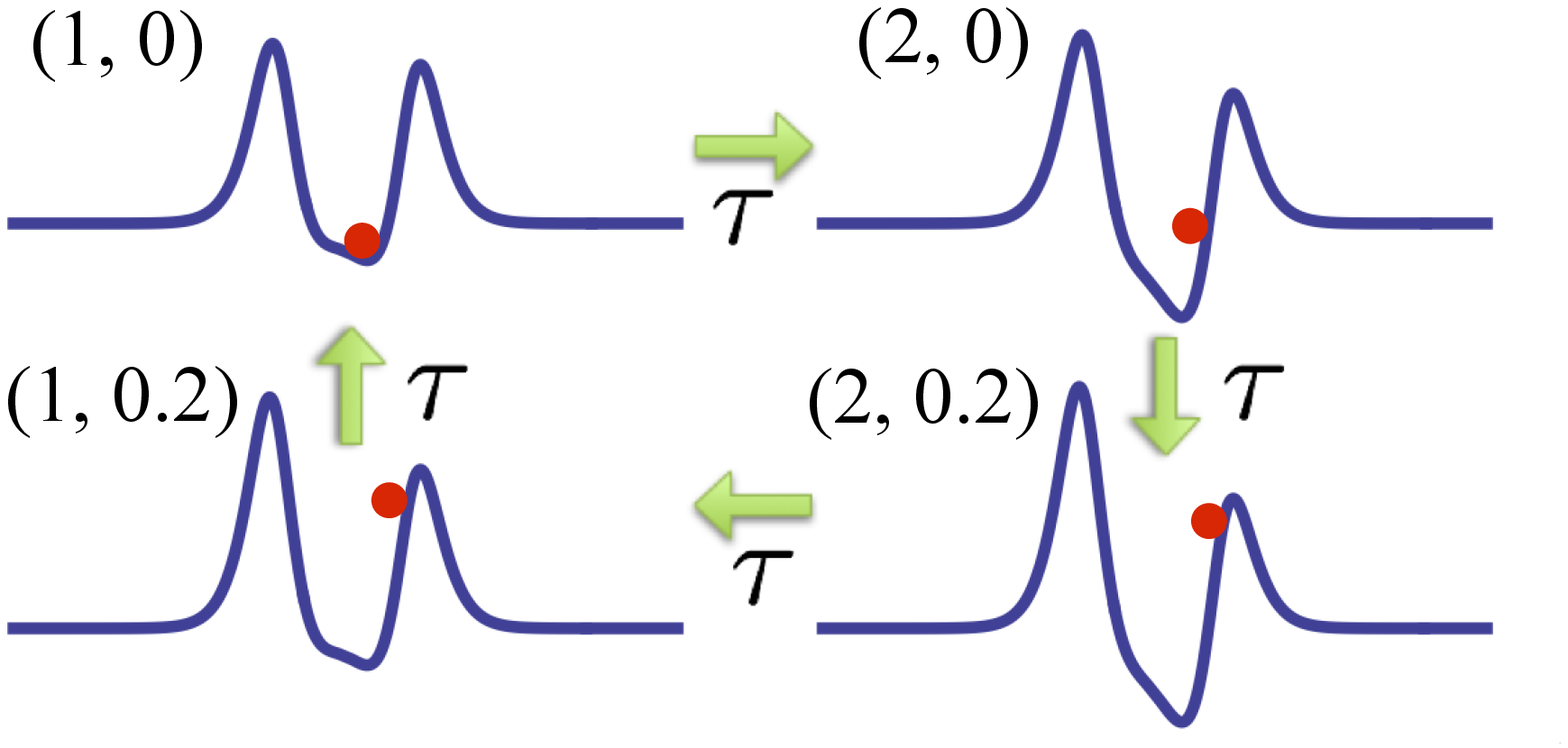}
\end{picture}

\begin{picture}(400,160)
\put(-45,100){\large (b)}
\includegraphics[scale=0.55]{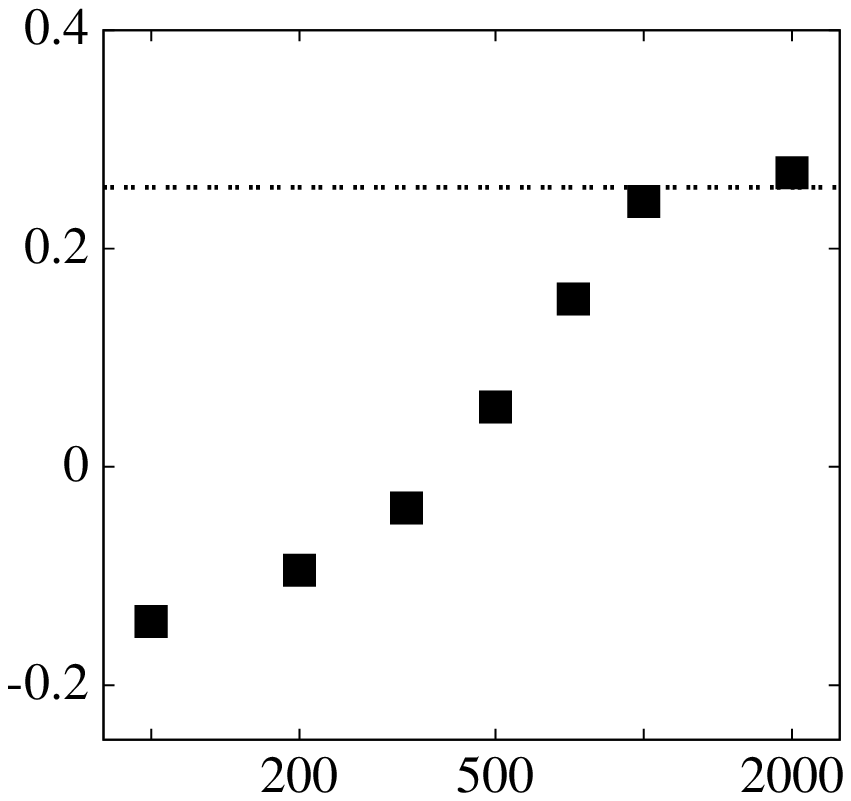}
\put(-45,-10){\Large $\tau$}
\put(-160,55){\rotatebox{90}{\large $\bbkt{Q}_\eq$}}
\end{picture}

\begin{picture}(10,25)
\put(-15,130){\large (c)}
{\includegraphics[scale=0.45]{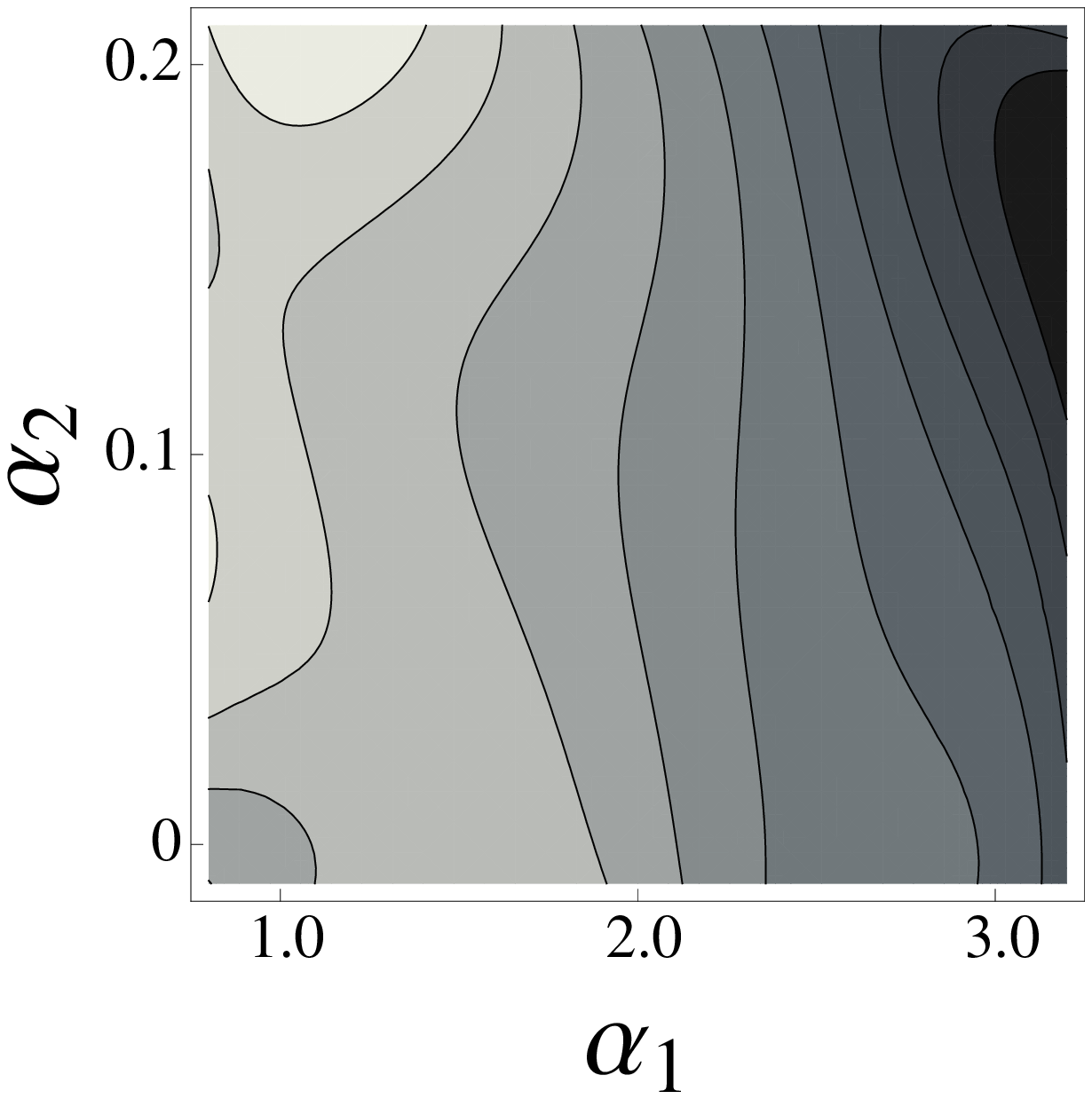}}
\end{picture}

\caption{
(Color online) (a) Ratchet potential
$V(x;\alpha_1,\alpha_2)= \{x^4-(x+\alpha_1)^2-5\}\{\tanh(x+3+\alpha_2)-\tanh(x-3)\}/100
$ and the applied cyclic protocol.
The values of $(\alpha_1, \alpha_2)$ are written in respective figures.
The particle (solid red circle) evolves according to Eq.~\eqref{e:Langevin},
where we take $\gamma=1$ and $k_\mathrm{B} T=0.3$. 
For the respective change (green arrow), 
either $\alpha_1$ or $\alpha_2$ is changed in a constant speed ($\propto \tau^{-1}$).
(b) Operation time $\tau$ vs total pumping $\sbkt{Q}_\eq$ observed in a cycle. 
We start from $(\alpha_1, \alpha_2)=(1,0)$ after preparing its steady state
and continue the operation without stopping until we return to $(1,0)$.
After finishing the operation, we continue calculation until the system reaches equilibrium.
The dashed line corresponds to the estimate $\sbkt{Q}_\eq=0.256$ explained in (c).
(c) The map of pumping density $\sbkt{{\cal J}}_{\rhoeq}$. 
$\rhost_{\varepsilon}(\Gamma)$ is calculated in the counterpart driven by
$\varepsilon=2/3 \times 10^{-2}$.
$\sbkt{Q}_\eq$ for the adiabatic limit of the protocol in (a) is estimated as $0.256$ from the integration of  $\sbkt{{\cal J}}_{\rhoeq}$ in Eq.~\eqref{e:pump3}.
The lightest shading corresponds to $0\le \sbkt{{\cal J}}_{\rhoeq} < 0.15$ and the black to $\sbkt{{\cal J}}_{\rhoeq} > 1.5$}
\label{fig:ratchet}
\end{figure}

\begin{figure}
\begin{picture}(190,150)
\put(-25,80){\large (a)}
{\includegraphics[scale=0.4]{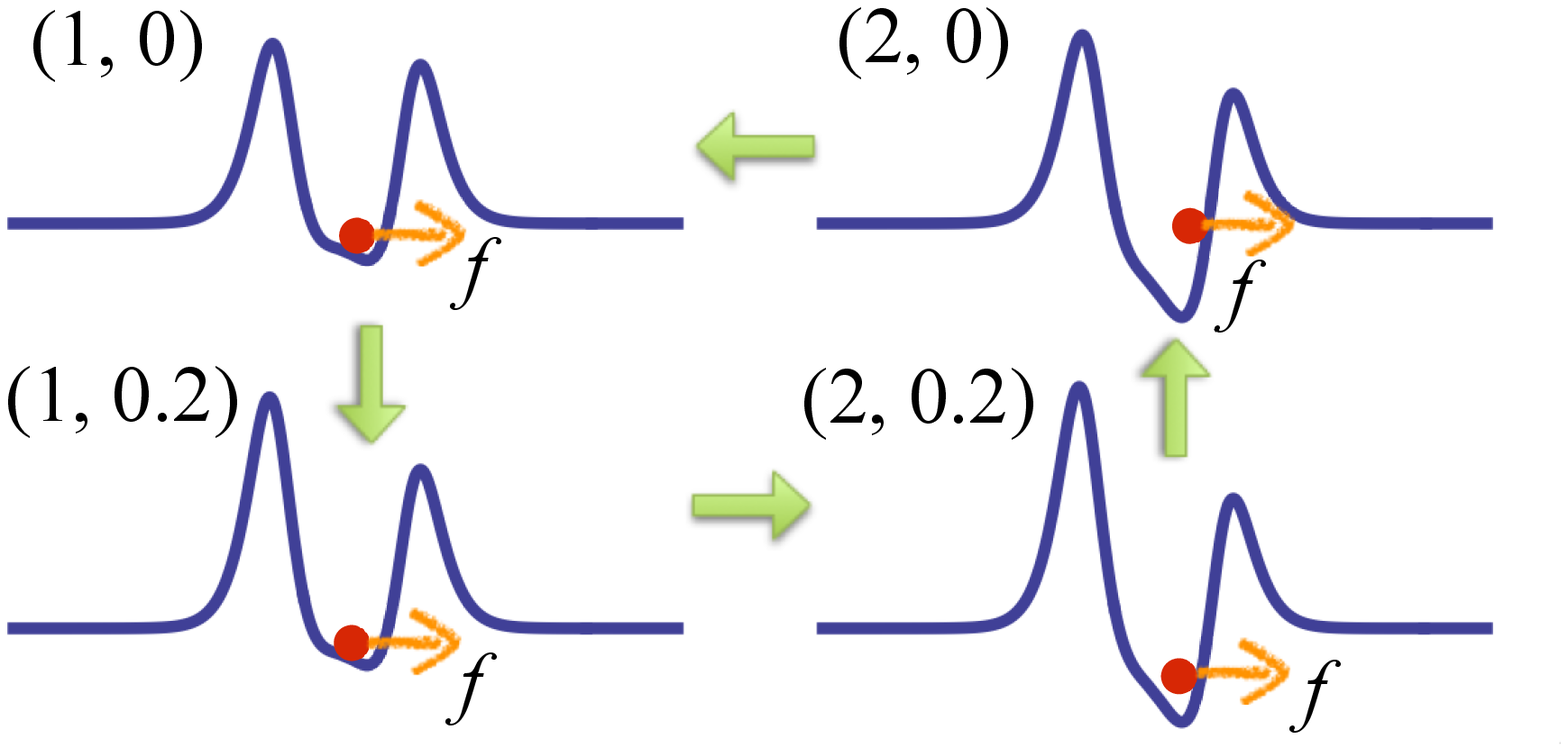}}
\end{picture}

\begin{picture}(160,160)
\put(-40,120){\large (b)}
{\includegraphics[scale=0.55]{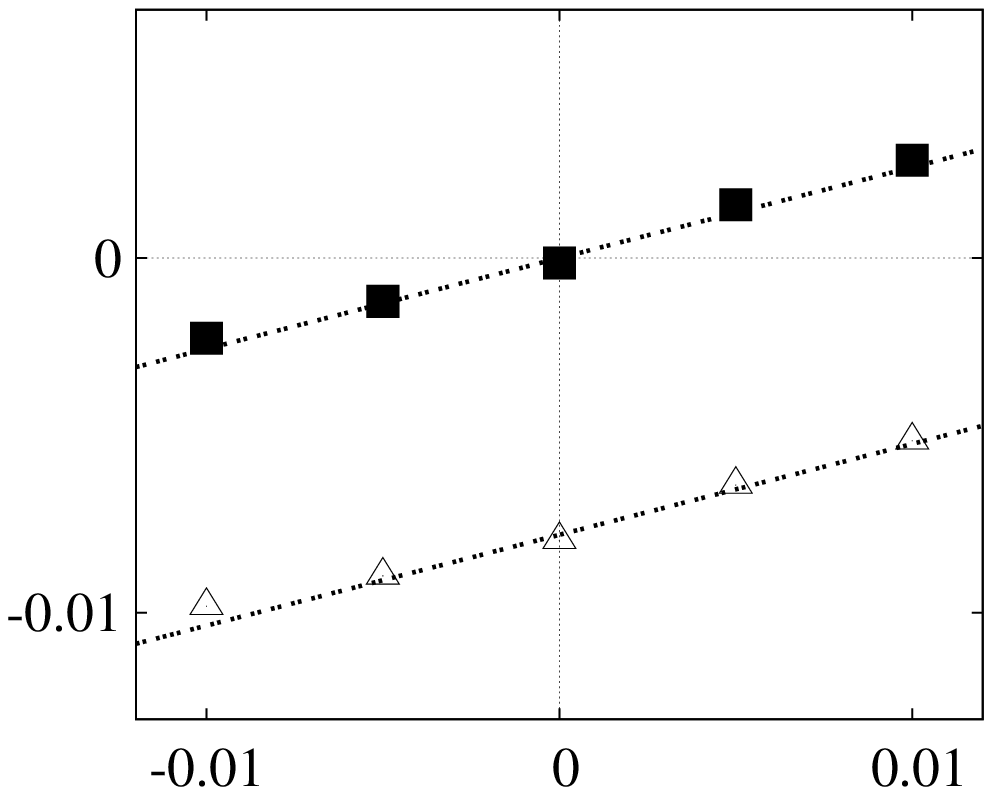}}
\put(-90,-20){\Large ${\varepsilon ~(=\beta f)}$}
\put(-130,95){\large ${-\log\bbkt{e^{-\beta W}}^{\dagger}_{\varepsilon}}$}
\put(-50,25){\large $\beta \sbkt{W}^{\dagger}_{\varepsilon}$}
\end{picture}

\vspace*{1cm}
\caption{
(Color online) 
(a) The reverse protocol applied to the system. The particle is always driven by $f$. 
(b) $\varepsilon$ vs $-\log\sbkt{e^{-\beta W}}^{\dagger}_{\varepsilon}$ (solid square) and $\beta \sbkt{W}_{\varepsilon}^{\dagger}$ (open triangle) for
$\tau=1000$. We start from the initial conditions in the steady state under the conjugate driving and start the reverse of the cyclic protocol in (a). 
We measure $W$ up to the end of the change for $\bm \alpha$
without calculating the relaxation process after the change.
The dashed lines are proportional to $\varepsilon$ with a slope $0.256$
whose value was estimated from the pumping density ${\cal J}(\Gamma)$ in Fig.~1~(c).
}\label{fig:ratchet-ness}
\end{figure}

We take a flashing ratchet model (see Fig.~\ref{fig:ratchet}).
The position of a particle evolves in a one-dimensional periodic potential $V(x;\alpha_1,\alpha_2)$ according to the Langevin equation
\begin{equation}
\gamma \dot x = -\frac{\partial V}{\partial x}+\sqrt{2\gamma k_\mathrm{B} T} \xi(t),
\label{e:Langevin}
\end{equation}
where $\gamma$ is the friction constant, $T$ is the temperature of the environment, and  $k_\mathrm{B}$ is the Boltzmann constant.
The ratchet potential $V(x;\alpha_1,\alpha_2)$ is operated externally by changing $(\alpha_1, \alpha_2)$ in the operation time $\tau$ [Fig.~\ref{fig:ratchet}(a)].
The total pumping $\sbkt{Q}_\mathrm{eq}$ in this example corresponds to the mean shift of the particle.

As shown in Fig.~\ref{fig:ratchet}(b),  $\sbkt{Q}_\mathrm{eq}$ converges to a certain value in larger values of $\tau$,
which will be  the value for the adiabatic limit.
Indeed, it is approximately equal to 
the expected total pumping in the adiabatic limit indicated by the dashed line, 
which is estimated from the calculation of the pumping density $\sbkt{{\cal J}}_{\rhoeq}$.
To determine $\sbkt{{\cal J}}_{\rhoeq}$, we calculate $\rhost_{\varepsilon}(\Gamma)$
for the system under the conjugate driving $\varepsilon=\beta f$,
\begin{equation}
\gamma \dot x = -\frac{\partial V}{\partial x}+f+\sqrt{2\gamma k_\mathrm{B} T} \xi(t)
\label{e:Langevin2}
\end{equation}
for a certain $\bm \alpha$,
from which we determine ${\cal J}(\Gamma)$ in Eq.~\eqref{e:density} 
and take the average of ${\cal J}(\Gamma)$ by $\rhoeq(\Gamma)$.
We repeat this procedure for various $\bm\alpha$,
and obtain the contour plot of $\sbkt{{\cal J}}_{\rhoeq}$ shown in Fig.~\ref{fig:ratchet}(c).

Next, we apply the reverse cyclic protocol [Fig.~\ref{fig:ratchet-ness}(a)]
to Eq.~\eqref{e:Langevin2} and calculate the work
$W=
\int_{\pathad} d{\bm\alpha} \cdot \nabla_{\alpha} V$.
From the ensemble of $W$ for a slow operation, 
we calculate $-\log\sbkt{
e^{-\beta W}
}^{\dagger}_{\varepsilon}$,
which is proportional to $\varepsilon$ as shown in Fig.~\ref{fig:ratchet-ness}(b).
The slope in the figure ($\tau=1000$) is close to the total pumping  $\sbkt{Q}_\eq$ for the adiabatic limit.
This coincidence corresponds to the convergence of $\sbkt{Q}_\eq$ around $\tau=1000$ [see Fig.~\ref{fig:ratchet}(b)].
In Fig.~\ref{fig:ratchet-ness}(b), we supplementarily plot $\beta \sbkt{W}^{\dagger}_{\varepsilon}$.
Since $-\log\sbkt{e^{-\beta W}}^{\dagger}_{\varepsilon} = \beta \sbkt{W}^{\dagger}_{\varepsilon}+O(\tau^{-2})$,
the line deviates from the origin at $\varepsilon=0$ due to the finiteness of $\tau$. 
However, the slope of $\beta \sbkt{W}^{\dagger}_{\varepsilon}$ is also close to $\sbkt{Q}_\eq$ for the adiabatic limit. 
When we take a smaller value of $\tau$, 
the slope of $\beta \sbkt{W}^{\dagger}_{\varepsilon}$ or $-\log\sbkt{e^{-\beta W}}^{\dagger}_{\varepsilon}$ 
becomes less steep consistently with the decrease of $\sbkt{Q}_\eq$.
This result may suggest that the slope of $\beta \sbkt{W}^{\dagger}_{\varepsilon}$ is an informative quantity for various pumping protocols with finite speed.

\subsection{Pumping densities in three state model}

We here study a simpler example which can be solved exactly.
We take a one-dimensional Markov jump model of three states ($x=1,2$ and $3$) with a periodic boundary condition identifying $x=3$ with $x=0$.
It acts as both a heat and a stochastic pump simultaneously.

In order to design the rate constants for the jump, 
we assume virtual energy barriers at every midpoint of the neighboring two states.
We set the energies of the three states as $v_1$, $v_2$ and $v_3$,
and the energies of the barriers as $u_{12}$, $u_{23}$ and $u_{31}$, respectively.
Then, we express the jump rates $R_{yx}$ from $x$ to $y$ 
as $R_{yx}=e^{-\beta (u_{yx}-v_x)}$.
We assume the parameters for the operation as $\bm\alpha=(v_2,u_{23})$.

\begin{figure}[t]
\centering
\begin{picture}(400,180)
\put(-20,100){(a)}
\includegraphics[scale=0.4]{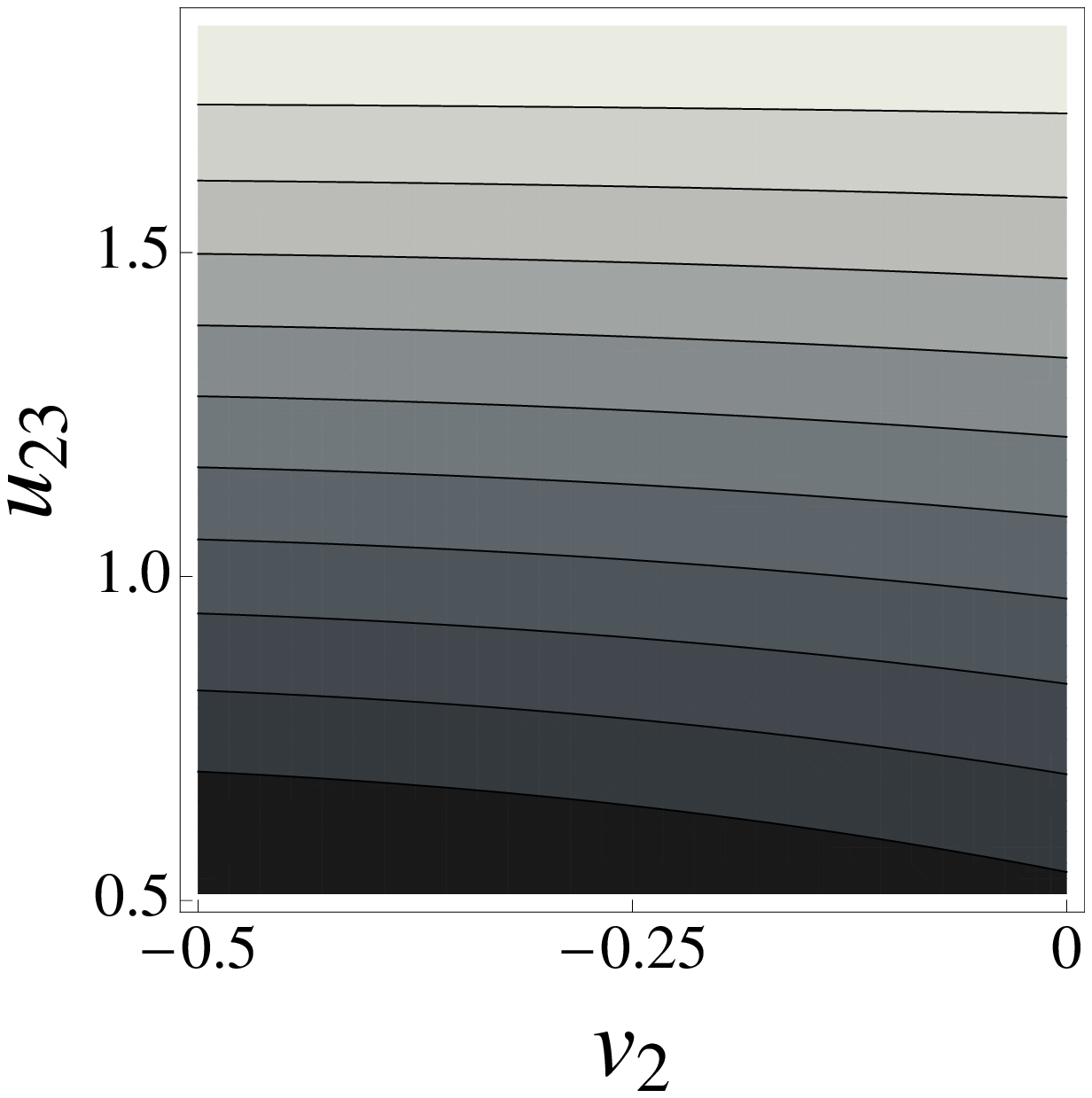}
\end{picture}

\begin{picture}(-30,0)
\put(-20,100){(b)}
{\includegraphics[scale=0.4]{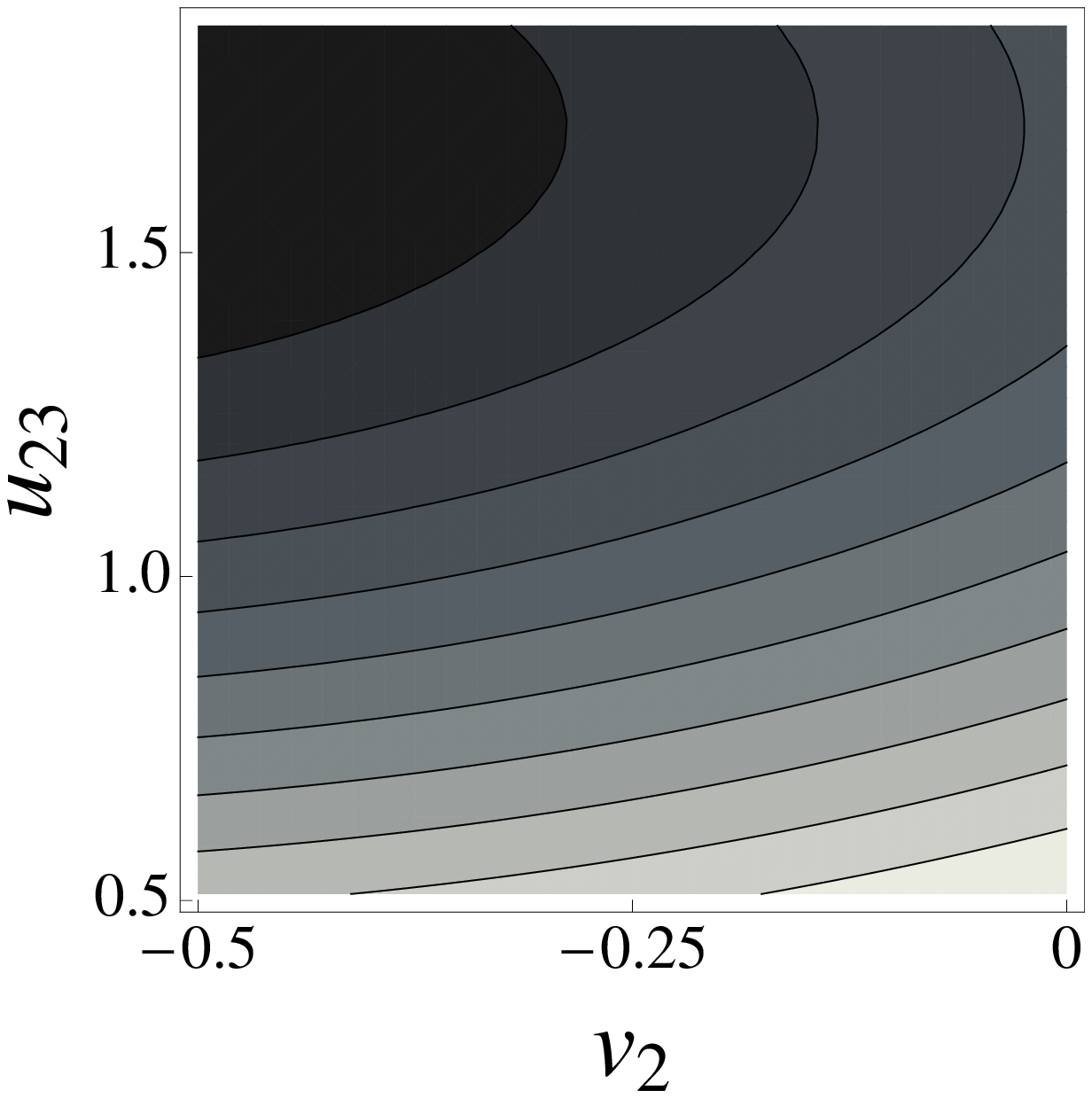}}
\end{picture}
\caption{
Contour plot in a grey scale for pumping density $|\sbkt{{\cal J}}_{\rhoeq}|$
determined by $\rhoeq(x)$ and $\rhost_{\varepsilon}(x)$,
where $\rhost_{\varepsilon}(x)$ is calculated 
from the rate matrices.
(a) Heat-pumping density calculated from the rate matrix \eqref{eq:rate1}.
The operational parameters are $v_2$ and $u_{23}$ and 
the other parameters are fixed as $v_1=v_3=0$ and $u_{12}=u_{31}=1$.
The lightest shading corresponds to $|\sbkt{{\cal J}}_{\rhoeq}|<0.011$ and the black to  $|\sbkt{{\cal J}}_{\rhoeq}|>0.099$.
(b) Stochastic pumping density calculated from the rate matrix \eqref{eq:rate2}.
The parameters are the same as in (a).
The lightest shading corresponds to $|\sbkt{{\cal J}}_{\rhoeq}|<0.099$ and the black to  $|\sbkt{{\cal J}}_{\rhoeq}|>0.135$.}
\label{fig:density}
\end{figure}

First, we show the pumping density when the system works as a heat pump.
For this purpose, we assume the system is in contact with two heat baths:
The one (say $\beta_1$) is in the region $1\le x < 2.5$ 
and the other (say $\beta_2$) is in $2.5\le x < 4 (=1)$.
The rate matrix for the conjugate driving $\varepsilon=\beta_2-\beta_1$
is expressed as
\begin{equation}
R^\varepsilon=\left(
\begin{array}{ccc}
-\lambda_1^\varepsilon & R_{12}e^{\frac{\varepsilon}{2} (u_{12}-v_1)} & R_{13}e^{-\frac{\varepsilon}{2} (u_{13}-v_1)} \\
R_{21}e^{\frac{\varepsilon}{2} (u_{12}-v_2)} & -\lambda_2^\varepsilon & R_{23}e^{-\frac{\varepsilon}{2} (u_{23}-v_2)}\\
R_{31}e^{-\frac{\varepsilon}{2} (u_{12}-v_3)} & R_{32}e^{\frac{\varepsilon}{2} (u_{23}-v_3)} & -\lambda_3^\varepsilon
\end{array}
\right)
,
\label{eq:rate1}
\end{equation}
where $\lambda_x^{\varepsilon}=\sum_{y\neq x} R_{xy}^\varepsilon$.
We numerically calculate 
the probability distribution $\rhost_{\varepsilon}(x)$
for various values of $\bm\alpha$. 
Figure \ref{fig:density}(a)
shows the pumping density resulting from the set of 
$\rhost_{0}(x)$ and $\rhost_{\varepsilon}(x)$.

Second, we map the pumping density when the same system works as a stochastic pump.
For the conjugate driving, we consider a uniform nonconservative force $f$ in the direction of $x$, i.e. $\varepsilon=\beta f$.
The rate constants $R_{yx}^\varepsilon$ are
\begin{equation}
R^\varepsilon=\left(
\begin{array}{ccc}
-\lambda_1^\varepsilon & R_{12}e^{-\frac{\varepsilon}{2}} & R_{13}e^{\frac{\varepsilon}{2}} \\
R_{21}e^{\frac{\varepsilon}{2}}& -\lambda_2^\varepsilon & R_{23}e^{-\frac{\varepsilon}{2}}\\
R_{31}e^{-\frac{\varepsilon}{2}} & R_{32}e^{\frac{\varepsilon}{2}} & -\lambda_3^\varepsilon
\end{array}
\right)
.
\label{eq:rate2}
\end{equation}
The pumping density is shown in Fig.~\ref{fig:density}(b).
These maps show that the system pumps both heat and particle simultaneously.

\section{Discussions}
We have developed a unified viewpoint on pumping and nonequilibrium thermodynamics
by introducing a driven counterpart to pumping.
With our unified viewpoint one can rederive various pumping results such as Eqs.~\eqref{e:pump1}, \eqref{e:pump1-2}, \eqref{e:density}, \eqref{e:work2} and \eqref{e:workG}.
From a theoretical point of view, the connection of total pumping $\sbkt{Q}_\eq$ to the stationary distribution $\rhost_{\varepsilon}$ in the driven counterpart or to the Fisher information matrix  \eqref{e:pump1} and \eqref{e:pump1-2}
is most interesting.
We expect that the accumulated knowledge on the Fisher information matrix 
provides a new viewpoint on pumping,
while it remains as a future work.

From a point of applicability,
we have related the total pumping $\sbkt{Q}_\eq$ in equilibrium 
to the work  $\sbkt{W}_{\varepsilon}$  or $-\log\sbkt{e^{-\beta W}}^{\dagger}_{\varepsilon}$ in the driven counterpart
as shown in Eqs.~\eqref{e:work2} and \eqref{e:workG}.
These relations are useful
when $Q$ is difficult to observe but $W$ is measurable.
As an example of application, we evaluate the pumping $\sbkt{Q}_\eq$ from $W$ in a mesoscopic pump. See Fig.\ref{fig:ratchet-ness}(b).

The work relation \eqref{e:work2} accompanied by Eqs.~\eqref{e:pump3} and \eqref{e:density}
shows that meso- or macroscopic force in NESS is no longer a potential force due to the geometric effects of pumping.
We need to use vector potential related to pumping in addition to the usual scalar potential.
We comment that the geometric effect of excess heat reported in \cite{Sagawa_Hayakawa}
has the same origin as 
the geometric effects of pumping  in Eq.~\eqref{e:density} 
and in \cite{Parrondo,Sinitsyn_Nemenman,Astumian,Ohkubo,Ren_Hanggi_Li}.
This is because relation \eqref{e:work2} is a version of an extended Clausius relation, which makes a connection between 
the excess heat and the entropy change \cite{KNST}.

\paragraph*{Acknowledgement}
The author is grateful to Hal Tasaki for stimulating discussions and a critical reading of the manuscript,
and to Keiji Saito for suggestions and comments, especially on the relation of \eqref{e:pump1} to the Fisher information matrix. 
This work was supported by JSPS/MEXT KAKENHI Grants No. 23540435 and No. 25103002.

\section{Appendix}

\renewcommand{\rhost}{\rho^\mathrm{st}}
\renewcommand{\rhoeq}{\rho^\mathrm{eq}}
\newcommand{\rhoeqO}{\rho_\mathrm{eq}}

In this Appendix, we use fully specified notations:
$\rhost_{\bm\alpha,\varepsilon}(\Gamma)$, $\rhoeq_{\bm\alpha}(\Gamma)$, $\psi_{\bm\alpha}^{\varepsilon}(\Gamma)$ and $\psi_{\bm\alpha}^\eq(\Gamma)$ 
 instead of  $\rho_{\varepsilon}(\Gamma)$, $\rhoeqO(\Gamma)$,
$\psi^{\varepsilon}(\Gamma)$ and $\psi^\eq(\Gamma)$.
The averages are
$\sbkt{f}^{\patha}_{\varepsilon}$ and $\bbkt{f}^{\bm\alpha}_{\rho_\varepsilon}$
instead of $\sbkt{f}_{\varepsilon}$ and $\bbkt{f}_{\rho_\varepsilon}$.

\subsection{Derivation of  Eq.~\eqref{e:pump1}}

We reported in \cite{KN-long} that the probability distribution of NESS
under the steady driving field $\varepsilon$ has a linear response representation,
\begin{eqnarray}
\rhost_{\bm\alpha,\varepsilon}(\Gamma)
=
\rhoeq_{\bm\alpha}(\Gamma)~
\exp{\left[-\varepsilon
\bbkt{Q}^{(\bm\alpha)}_{\Gamma^*\rightarrow\eq}
\right]}
+\oet
,
\label{e:KN}
\end{eqnarray}
where a conditioned expectation is defined as
\begin{equation} 
\sbkt{Q}_{\Gamma\rightarrow\eq}^{(\bm\alpha)}
=\int {\calD}\pathG\,\delta(\Gamma(\ti)-\Gamma)\WT_{(\bm\alpha)}^\mathrm{eq}[\pathG]\,Q(\pathG), 
\end{equation}
with a fixed initial state $\Gamma$.
The notation $(\bm\alpha)$ represents the protocol in which the parameters are kept constant at $\bm\alpha$ and $\WT_{(\bm\alpha)}^\mathrm{eq}=\WT_{(\bm\alpha),0}$.
The conditioned average $\sbkt{Q}^{(\bm\alpha)}_{\Gamma\rightarrow\eq}$
gives the total transfer observed in the relaxation process
from the state $\Gamma$.
There is no transfer on average in equilibrium,
i.e.,
\begin{equation}
\int d\Gamma \rhoeq_{\bm\alpha}(\Gamma) \bbkt{Q}_{\Gamma\rightarrow\eq}^{(\bm\alpha)}=0
.
\label{e:eq}
\end{equation}

We first concentrate on
the protocol of an infinitesimal stepwise change from
$\bm\alpha$ to $\bm{\alpha}'=\bm{\alpha}+\Da$. 
Even though we do not observe any current before the stepwise change, 
we may observe it in the relaxation process after the stepwise change.
Noting that $\sbkt{Q}_{\Gamma\rightarrow\eq}^{(\bm{\alpha}')}$ is the total transfer in the relaxation from the state $\Gamma$,
the total transfer after the stepwise change is
\begin{eqnarray}
\bbkt{Q}^{\patha}_\eq
&=&
\int d\Gamma  \rhoeq_{\bm\alpha}(\Gamma) \bbkt{Q}_{\Gamma\rightarrow\eq}^{(\bm{\alpha}')},\nonumber\\
&=&
-\int d\Gamma  \left(
\rhoeq_{\bm{\alpha}'}(\Gamma)
-
\rhoeq_{\bm\alpha}(\Gamma)
\right)
\bbkt{Q}_{\Gamma\rightarrow\eq}^{(\bm{\alpha}')}
,
\label{e:eq2}
\end{eqnarray}
where we subtract Eq.~\eqref{e:eq} from the first line of Eq.~\eqref{e:eq2}
in order to obtain the expression in the second line.
If $\rhoeq_{\bm{\alpha}'}(\Gamma)
=
\rhoeq_{\bm\alpha}(\Gamma)+O(|\Da|)$ and
$\sbkt{Q}_{\Gamma\rightarrow\eq}^{(\bm{\alpha}')}=\sbkt{Q}_{\Gamma\rightarrow\eq}^{(\bm\alpha)}+O(|\Da|)$,
then
\begin{eqnarray}
\bbkt{Q}^{\patha}_\eq
&=&
-{\Da}\cdot\int d\Gamma  (\nabla_{\alpha}\rhoeq_{\bm\alpha}(\Gamma)) 
\bbkt{Q}_{\Gamma\rightarrow\eq}^{(\bm\alpha)}
\label{e:dv1}
\end{eqnarray}
with an error of $O(|\Da|^2)$.

As the next step, we refer to the representation \eqref{e:KN},
where $\sbkt{Q}_{\Gamma\rightarrow\eq}^{(\bm{\alpha})}$ is related to
$\rhost_{\bm\alpha,\varepsilon}$.
Therefore, it is apparent that 
\begin{equation}
\bbkt{Q}_{\Gamma\rightarrow\eq}^{(\bm{\alpha})}
=
\left.
\partial_\varepsilon\psi_{\bm\alpha}^{\varepsilon}(\Gamma^*)
\right|_{\varepsilon=0}
.
\label{e:pdf-J}
\end{equation}
Substituting Eq.~\eqref{e:pdf-J} into Eq.~\eqref{e:dv1},
we have
\begin{eqnarray}
\bbkt{Q}^{\patha}_\eq
&=&
-{\Da}\cdot\int d\Gamma  (\nabla_{\alpha}\rhoeq_{\bm\alpha}(\Gamma)) 
\left.
\partial_\varepsilon\psi_{\bm\alpha}^{\varepsilon}(\Gamma)
\right|_{\varepsilon=0}
,\nonumber\\
&=&
{\Da}\cdot
\int d\Gamma  \rhoeq_{\bm\alpha}(\Gamma)
\left.
\nabla_{\alpha}\psi_{\bm\alpha}^\eq(\Gamma)
\partial_\varepsilon \psi_{\bm\alpha}^{\varepsilon}(\Gamma)
\right|_{\varepsilon=0}
,
\label{e:dv1-2}
\end{eqnarray}
where the negligible error term of $O(|\Da|^2)$ is ignored.
We used $\rhoeq_{\bm\alpha}(\Gamma)=\rhoeq_{\bm\alpha}(\Gamma^*)$
to obtain the first line of Eq.~\eqref{e:dv1-2}.

Finally, we note that any adiabatic protocol is 
the accumulation of infinitesimal steps.
We need to extend Eq.~\eqref{e:dv1-2} to
the line integral along the protocol $\patha$,
as is expressed in Eq.~\eqref{e:pump1}.

In order to arrive at expression \eqref{e:pump1-2}, 
we use an identity,
\begin{equation}
\int d\Gamma \rho(\Gamma)\frac{\partial^2 \psi(\Gamma)}{\partial \alpha\partial\varepsilon}
=\int d\Gamma \rho(\Gamma)\frac{\partial \psi(\Gamma)}{\partial \alpha}\frac{\partial \psi(\Gamma)}{\partial \varepsilon},
\end{equation}
which is derived from integration by parts and the conservation law $\int \rho(\Gamma) d\Gamma=1$.

\subsection{Derivation of  Eq.~\eqref{e:work2}}

We start from Eq.~\eqref{e:pump1-2}.
Substituting Eq.~\eqref{e:pdf-J} into Eq.~\eqref{e:pump1-2}, we have
\begin{eqnarray}
\bbkt{Q}^{\patha}_\eq
&=&
\int_{\patha} d{\bm \alpha}\cdot
\int d\Gamma \rhoeq_{\bm\alpha}(\Gamma) \nabla_{\alpha} \bbkt{Q}_{\Gamma^*\rightarrow\eq}^{(\bm{\alpha})}
\nonumber\\
&=&
\int_{\patha} d{\bm \alpha}\cdot
\int d\Gamma \rhoeq_{\bm\alpha}(\Gamma) \bbkt{Q}_{\Gamma^*\rightarrow\eq}^{(\bm{\alpha})}
\nabla_{\alpha} \psi_{\bm\alpha}^{\eq}(\Gamma),
\label{e:dv2}
\end{eqnarray}
where we applied the integration by parts.
As the expression \eqref{e:KN}  leads to
\begin{equation}
\rhoeq_{\bm\alpha}(\Gamma)~\bbkt{Q}^{(\bm\alpha)}_{\Gamma^*\rightarrow\eq}
=
-\frac{\rhost_{\bm\alpha,\varepsilon}(\Gamma)
-
\rhoeq_{\bm\alpha}(\Gamma)}{\varepsilon}
+O(\varepsilon),
\end{equation}
Eq.~\eqref{e:dv2} is transformed as
\begin{eqnarray}
\bbkt{Q}^{\patha}_\eq
&=&
-\frac{1}{\varepsilon}
\int_{\patha} d{\bm \alpha}\cdot
\int d\Gamma 
[\rhost_{\bm\alpha,\varepsilon}(\Gamma)
-
\rhoeq_{\bm\alpha}(\Gamma)]
\nabla_{\alpha} \psi_{\bm\alpha}^{\eq}(\Gamma)
\nonumber\\
&=&
\frac{\beta}{\varepsilon}
\int_{\patha} d{\bm \alpha}\cdot
\int d\Gamma 
\rhost_{\bm\alpha,\varepsilon}(\Gamma)
\left(\nabla_{\alpha} F -\nabla_{\alpha} H(\Gamma) \right)
,
\end{eqnarray}
where $F$ is the equilibrium free energy satisfying $\sbkt{\nabla_{\alpha}H}^{\bm\alpha}_{\rhoeq}=\nabla_{\alpha}F$.
Thus, we arrive at the final formula 
\begin{equation}
\bbkt{Q}^{\patha}_\eq 
=
-{\beta}\frac{\bbkt{W}^{\patha}_{\varepsilon}-\Di F }{\varepsilon}
+ O(\varepsilon),
\label{e:work1}
\end{equation}
where $\Di F = \int_{\patha} d{\bm \alpha}\cdot  \nabla_{\alpha}F$
and 
$
\sbkt{W}^{\patha}_{\varepsilon}=
\int_{\patha} d{\bm \alpha}\cdot 
\int d\Gamma \rhost_{\bm\alpha,\varepsilon}(\Gamma)\nabla_{\alpha}H(\Gamma)
$.
Since $\Di F=0$ in cyclic protocols, we have Eq.~\eqref{e:work2} as a direct consequence of Eq.~\eqref{e:work1}.


\end{document}